\documentstyle[preprint,aps,floats]{revtex}

\begin{document}

\vskip 1.5 cm

\preprint{\tighten \vbox{\hbox{} }}

\title{$K_{L,S} \to \pi \pi \nu \bar{\nu}$ Decays Within and 
Beyond the Standard Model}

\author{Cheng-Wei Chiang\footnote{Electronic address:
chengwei@andrew.cmu.edu} 
and Frederick J. Gilman\footnote{Electronic address: 
gilman@cmuhep2.phys.cmu.edu}}

\address{
Department of Physics,
Carnegie Mellon University,
Pittsburgh, Pennsylvania 15213}

\maketitle

{\tighten
\begin{abstract}
The decays $K_{L,S} \to \pi^o \pi^o \nu \bar{\nu}$ and 
$K_{L,S} \to \pi^+ \pi^- \nu \bar{\nu}$ 
involve the weak transition from a strange to a non-strange quark. 
Although they have considerably smaller branching ratios than 
those for the corresponding rare processes involving single pions, 
some may be more distinguishable experimentally from background 
processes and thus could provide another probe of the 
$s \to d \nu \bar{\nu}$ transition.  Using recent knowledge of 
the CKM matrix elements and measurements of related processes, 
we give improved predictions for their branching ratios both 
within and beyond the Standard Model.
\end{abstract}
}

\newpage

\section{Introduction}

Historically, rare kaon decays have provided a crucial testing 
ground in which to study flavor-changing neutral current (FCNC) 
and CP-violating phenomena. Within the context of the Standard 
Model, the observation of such decays gives us further information 
on the Cabibbo-Kobayashi-Maskawa (CKM) matrix elements and provides 
independent checks on the consistency of our understanding 
drawn from other measurements, such as those of semileptonic decays 
and B-meson decays.  They also serve as a good place to look for 
physics beyond the Standard Model.  Even if new physics is first 
observed outside the kaon sector, we will want to know the footprint 
it leaves in rare kaon decays and specifically on the 
$s \to d$ weak transition. 

In this paper we focus on the decays $K_{L,S} \to \pi^o \pi^o 
\nu \bar{\nu}$ and $K_{L,S} \to \pi^+ \pi^- \nu \bar{\nu}$.  
As noted previously\cite{GHL94,LV96}, 
at the quark level these FCNC processes involve the 
$s \to d \nu \bar{\nu}$ transition.  In the Standard Model, 
this transition is dominated by short-distance contributions 
involving loop diagrams that contain W and Z bosons and heavy 
quarks\cite{IL81}. With the top quark mass much greater than 
that of the charm quark, the imaginary part of the 
amplitude for this transition arises almost entirely from 
loop diagrams with top quarks and the resulting amplitude 
is then proportional to the CKM factor Im(${V_{td}}^* V_{ts}$).   

The decays, $K^+ \to \pi^+ \nu \bar{\nu}$ and 
$K_L \to \pi^o \nu \bar{\nu}$ are both governed by the same 
$s \to d \nu \bar{\nu}$ transition. The latter decay is 
CP-violating and dominated to high accuracy by the 
short-distance contribution.  As noted above, its amplitude 
is consequently proportional to Im(${V_{td}}^* V_{ts}$) in 
the Standard Model.  For both decays, the hadronic matrix 
elements of the relevant weak current can be related to ones 
which enter charged current semileptonic decays, whose direct 
measurement bypasses any theoretical uncertainty in the hadronic 
matrix element as well.  The branching ratios that are predicted 
in the Standard Model lie between roughly $10^{-11}$ and $10^{-10}$. 
With negligible long-distance contributions and little hadronic 
uncertainty, these decays have been pointed to as crucial for 
precision experimental tests of the Standard Model and correspondingly 
as places to look for new physics if the Standard Model 
fails\cite{B99,I99}.
Because of this, they are being pursued experimentally in 
spite of very difficult experimental backgrounds.

The rare decays $K_{L,S} \to \pi \pi \nu \bar{\nu}$ that we 
study in this paper have the same advantage of allowing a 
theoretical clean study of the $s \to d \nu \bar{\nu}$ transition, 
although some of the relevant semileptonic matrix 
elements are not as accurately measured experimentally.  As shown 
previously\cite{GHL94,LV96} in the Standard Model, they unfortunately 
have the serious disadvantage that their predicted branching 
ratios are several orders of magnitude smaller than for the decays 
involving a single pion.  It is nevertheless interesting to pursue 
them because they involve different combinations of the 
CP-conserving and CP-violating parts of the $s \to d \nu \bar{\nu}$ 
transition, and they thus provide additional handles on both the 
real and imaginary parts of amplitude.  Furthermore, from an 
experimental point of view, some or all of these decays may prove 
to be more susceptible to the extraction of a signal from the 
background. 

In this paper, we significantly refine the predictions for 
$K \to \pi \pi \nu \bar\nu$ decays in the Standard Model using 
recent knowledge of the CKM matrix elements and measurements of 
decay rates for related processes.  We also examine how large 
these branching ratios could be for physics beyond the Standard 
Model and find that there are significant, model-independent 
limits from other measurements.  The paper is organized as follows:
In Section 2, we provide the general framework for studying the 
$s \to d \nu \bar{\nu}$ transition and the contributions to it both 
within and beyond the Standard Model.  Section 3 sets forth its 
relationship to the $K \to \pi \pi \nu \bar\nu$ decays under discussion, 
the CP properties of the amplitudes involved in those decays, and 
the connection of the relevant hadronic matrix elements to measured 
semileptonic decays.  Numerical results are given in Section 4, 
followed by some conclusions in Section 5.

\section{General Framework}

The effective Hamiltonian for $s \to d \nu \bar\nu$ transitions 
takes the form
\begin{equation}
{\cal H} = \frac{G_F}{\sqrt{2}}\frac{\alpha}{2\pi \sin^2\theta_W}
  W_{ds} \left[ \left( {\bar s} \gamma_{\mu} (1 - \gamma_5 ) d \right) 
\left( {\bar \nu} \gamma^{\mu} (1 - \gamma_5 )\nu \right) \right] +~h.c.~,
\end{equation}
where the short-distance physics is lumped in $W_{ds}$.  In the Standard 
Model, one-loop contributions to $W_{ds}$ are dominated by penguin and 
box diagrams with intermediate charm and top quarks:  
\begin{equation}
W_{ds}^{SM}= \lambda_{sd}^c X(x_c ) + \lambda_{sd}^t X(x_t ) ~,
\end{equation}
where $\lambda_{sd}^i \equiv V_{is}^* ~V_{id}$, with $V_{ij}$ 
the appropriate CKM matrix element, and $x_i = {\bar{m}_i}^2 /{M_W}^2$.  
The QCD corrections to the short-distance contributions $X(x_i )$ 
have been calculated some time ago in leading order\cite{HL89,DDG} 
and then in next-to-leading order\cite{BB93,BB99}.  Since the top-quark 
mass is comparable to the weak scale, these corrections are very 
small for $X(x_t )$, as can be seen explicitly in the values 
given\cite{BB93,BB99} for $X(x_t)$ when written as 
$X(x_t) = \eta_t X_0 (x_t)$, with the 
QCD-uncorrected top quark contribution\cite{IL81} 
\begin{equation}
X_0(x_t) = \frac{x_t}{8}
\left[\frac{x_t + 2}{x_t - 1} + 
\frac{3x_t - 6}{(x_t - 1)^2}\log (x_t ) \right] ~,
\end{equation}
and the QCD correction factor $\eta_t = 0.994$. On the other hand, 
these corrections have considerable importance for $X(x_c )$.

The quantity $X(x_t )$ is roughly three orders of magnitude 
larger than $X(x_c )$, and since ${\mathrm Im}\lambda_{sd}^c = - 
{\mathrm Im}\lambda_{sd}^t$, the top contribution completely dominates 
in the imaginary part of $W_{ds}^{SM}$.  However, 
Re$\lambda_{sd}^t <<$ Re$\lambda_{sd}^c$, allowing the 
charm contribution, although still smaller in magnitude than 
that from top, to be roughly comparable and to interfere 
constructively in the real part of $W_{ds}^{SM}$.

As illustrative examples of physics that lies beyond the Standard 
Model, we consider two very different possibilities:

\begin{itemize}

\item{\underline{Effective Flavor-Changing Neutral Current (FCNC) 
interaction}. Such an interaction, as formulated by Nir and 
Silverman \cite{NS90,S92}, takes the form of an extra term in 
the effective Lagrangian of the form:
\begin{equation}
{\cal L}^{(Z)} = -~{g \over 4\cos\theta_W } U_{ds}~ 
\bar{d} \gamma_\mu (1 - \gamma_5 ) s ~Z^\mu ~.
\end{equation}
When combined with the coupling of the Z boson to neutrino-antineutrino 
pairs, one finds that
\begin{equation}
W_{ds}^{NP} = \frac{\pi^2}{\sqrt{2} G_F M_W^2} U_{ds} = 
0.93 \times 10^2 ~U_{ds}
\end{equation}
as the new piece of $W_{ds}$ in the effective 
Hamiltonian that corresponds to the basic process, 
$s \to d \nu \bar\nu$. 

Upper bounds for $U_{ds}$ have been determined by other 
processes involving K mesons and were summarized\cite{N99} 
recently to be
\begin{eqnarray}
|{\mathrm Re}(U_{ds})| &\leq&\ 10^{-5}, \\
|U_{ds}| &\leq&\  3\times 10^{-5}, \\
|{\mathrm Re}(U_{ds}) \, {\mathrm Im}(U_{ds})| 
&\leq&\ 1.3 \times 10^{-9}, \\
|{\mathrm Im}(U_{ds})| &\leq&\ 10^{-5}.
\end{eqnarray}
The bound on $|U_{ds}|$ arises from the decay 
$K^+ \to \pi^+ \nu \bar{\nu}$, whose width is proportional 
to $|W_{ds}|^2$.  It can be improved by using the most recent 
measurement\cite{Adler00}, of the branching ratio,  
BR($K^+ \to \pi^+ \nu \bar{\nu}) = 1.5 {}^{+3.4}_{-1.2} \times 10^{-10}$.  
This value is consistent with what is expected in the Standard Model 
and corresponds to $|W_{ds}| = 0.98 {}^{+0.80}_{-0.54} \times 10^{-3}$.
If we were to assume that the total branching ratio were due to 
new physics arising from $U_{ds}$, then the bound on $|U_{ds}|$ would 
be reduced from that in Eq. (7) to $|U_{ds}| < 1.6 \times 10^{-5}$.}

\item{\underline{Supersymmetry}.  A dominant supersymmetric 
effect arises from penguin diagrams involving  charged-Higgs 
plus top-quark intermediate states or squark and chargino 
intermediate states. These give additional pieces to 
the effective Hamiltonian of the form\cite{CI98}
\begin{equation}
W_{ds}^{NP} = \lambda_{sd}^t \frac{m_H^2}{M_W^2 \tan^2\beta} H(x_{tH}) 
+ \frac1{96}{ {\tilde\lambda}_t } ~,
\end{equation}
where $\tan\beta$ is the ratio of the two Higgs vacuum expectation 
values and $x_{tH} = {m_t}^2 /{M_{H^\pm}}^2$. The quantity $H(x)$ 
is given by 
\begin{equation}
H(x) = \frac{x^2}{8} 
\left[ -\frac{\log x}{(x-1)^2}+\frac1{x-1} \right] ~,
\end{equation}
The parameter ${\tilde\lambda_t}$ can be bounded by similar 
considerations to those that were used for $U_{ds}$.  The 
observed branching ratios for the decays $K_L \to \mu^+ \mu^-$ and
$K^+ \to \pi^+ \nu \bar\nu$ have been used to set the limits\cite{CI98} 
\begin{eqnarray}
|{\mathrm Re}{\tilde\lambda}_t| &\leq& 0.21 ~, \\
|{\tilde\lambda}_t| &\leq& 0.35 ~.
\end{eqnarray}
The most recent branching ratio for $K^+ \to \pi^+ \nu \bar\nu$ 
could be used to revise the last limit to $|{\tilde\lambda}_t |< 0.16$ .}

\end{itemize}

As we will see shortly, the limitations imposed by experiments 
on the parameters of both these examples of physics beyond the 
Standard Model lead to similar restrictions on how large the 
branching ratios can be for the processes we are studying.

\section{{\mbox{\boldmath $K \to \pi \pi \nu \bar\nu$}} Decays}

When the effective four-fermion operator relevant for the decay 
we are considering is sandwiched between the initial and final states, 
it factorizes into a product of matrix elements of the hadronic 
current and the leptonic current.  We will use isotopic spin to 
relate the hadronic matrix elements relevant to 
$K_{L,S} \to \pi \pi \nu \bar{\nu}$ to those for 
$K^+ \to \pi \pi e^+ \nu$, where the corresponding branching ratios 
(and hence squares of matrix elements) have been measured. 

We consider first the process $K_L \to \pi^o \pi^o \nu \bar{\nu}$. 
The $\pi^o \pi^o$ pair forms a CP-even state, and has total 
isospin, $I = 0$.  The $\nu \bar{\nu}$ pair, created by 
a virtual $Z^o$, is CP even as well.  Since 
there must be one unit of orbital angular momentum to allow the 
total angular momentum of the final state to be that of the 
initial $K_L$, namely zero, the final state is CP-odd. The overall 
decay process is then CP-conserving for the major (CP-odd) piece 
of the $K_L$, and the resulting amplitude is proportional to the 
real part of $W_{ds}$.  The opposite result holds for the 
piece of the $K_L$ that is proportional to $\epsilon$ and CP-even; 
the corresponding decay process is CP-violating and the amplitude 
is proportional to the imaginary part of $W_{ds}$.  It is thus 
suppressed on two counts and is completely negligible.

Using the relationship,
\begin{equation}
\langle \pi^o \pi^o \mid (\bar{s} d)_{\rm V-A} \mid K^o \rangle = 
\langle \pi^o \pi^o \mid (\bar{s} u)_{\rm V-A} \mid K^+ \rangle ~,  
\end{equation} 
we find that 
\begin{equation}
BR(K_L \to \pi^o \pi^o \nu \bar{\nu}) = 
{3 \alpha^2 |{\rm Re}W_{ds}|^2
\over 2 \pi^2 \sin^4 \theta_W |V_{us}|^2 } 
{\tau_{K_L} \over \tau_{K^+} } 
 BR(K^+ \to \pi^o \pi^o e^+ \nu ) ~,
\end{equation}
where the factor of 3 accounts for the three species of neutrinos.   

By relating the desired branching ratio to a measured one, we have 
avoided having either to do a calculation of the hadronic matrix 
elements or to perform a detailed analysis in terms of invariant 
amplitudes, as was done in previous analyses\cite{GHL94,LV96}. 
Of course, the final results for the branching ratio must be consistent, 
since both approaches agree with the available data on charged-current 
semileptonic decays, and in particular those for the decay rate for 
the process $K^+ \to \pi^o \pi^o e^+ \nu$. 
For the purposes of this paper of discussing the absolute 
and relative size of the various branching ratios within and beyond 
the Standard Model, it is considerably easier to formulate the 
results directly in terms of relationships to branching ratios for 
measured semileptonic decays.  
 
A similar formula can be obtained for $BR(K_S \to \pi^0
\pi^0 \nu \bar\nu)$, but with ${\mathrm Re}W_{ds}$ replaced by 
${\mathrm Im}W_{ds}$ and $\tau_{K_L}$ replaced by $\tau_{K_S}$.  
Since the $K_S$ has a much shorter lifetime and the major part 
of the $K_S$ corresponds to a transition that is CP-violating,   
this branching ratio is orders of magnitude smaller than that 
for $K_L \to \pi^o \pi^o \nu \bar{\nu}$.  
Although the part of the $K_S$ state proportional to $\epsilon$ 
corresponds a CP-conserving transition, the smallness of 
$\epsilon$ still gives rise to a net decay amplitude that is 
much smaller than that for the CP-even part of the $K_S$.

The analysis for the decay $K_L \to \pi^+ \pi^- \nu \bar\nu$ 
can be carried out analogously.  It is convenient to break it 
up into the cases where the $\pi^+ \pi^-$ pair in the final state 
has total isospin zero and one, since there is no interference 
between them in the decay rate.  For the isospin zero case, 
the argument about the CP properties of the final state is 
the same as given before, and we simply have a factor of two 
in the rate for the $\pi^+ \pi^-$ final state compared to that 
for the $\pi^o \pi^o$ final state discussed above:
\begin{equation}
BR(K_L \to (\pi^+ \pi^-)_{\rm I = 0} \nu \bar{\nu}) = 
{3 \alpha^2 |{\rm Re}W_{ds}|^2
\over \pi^2 \sin^4 \theta_W |V_{us}|^2 } 
{\tau_{K_L} \over \tau_{K^+} }
 BR(K^+ \to \pi^o \pi^o e^+ \nu ).
\end{equation}

This situation is slightly more complicated for the case where 
the $\pi^+ \pi^-$ pair has isospin, $I = 1$.  The $\pi \pi$ 
pair is still CP-even, but it must be in a p-wave.  There 
are two possible ways in which the total angular momentum of the 
final state can be zero, which correspond to the relative orbital 
angular momentum of the $\pi \pi$ and $\nu \bar{\nu}$ pairs being 
zero or one. As is expected when there is such limited phase space, 
the latter amplitude is strongly suppressed by centrifugal barrier 
effects compared to the former\cite{LV96}.  So we are left with a 
single amplitude where the relative orbital angular momentum is zero. 
The CP of the final state is even and the transition involving 
the major part of the $K_L$ is CP-violating.  Using the relationship,
\begin{equation}
\sqrt{2}~\langle (\pi^+ \pi^-)_{\rm I = 1} 
\mid (\bar{s} d)_{\rm V-A} \mid K^o \rangle = 
\langle (\pi^- \pi^o)_{\rm I = 1} 
\mid (\bar{s} u)_{\rm V-A} \mid K^o \rangle ~,  
\end{equation} 
we find that 
\begin{equation}
BR(K_L \to (\pi^+ \pi^-)_{\rm I = 1} \nu \bar{\nu}) = 
{3 \alpha^2 |{\rm Im}W_{ds}|^2
\over 4 \pi^2 \sin^4 \theta_W |V_{us}|^2 }  
BR(K_L \to \pi^\mp \pi^o e^\pm \nu ) ~.
\end{equation}
The total branching ratio for $K_L \to \pi^+ \pi^- \nu \bar\nu$ 
is then found by simply adding the two results above: 
$BR(K_L \to \pi^+ \pi^- \nu \bar{\nu}) = 
BR(K_L \to (\pi^+ \pi^-)_{\rm I = 0} \nu \bar{\nu})
+ BR(K_L \to (\pi^+ \pi^-)_{\rm I = 1} \nu \bar{\nu})$.

The corresponding formula for $K_S \to \pi^+ \pi^- \nu \bar\nu$
can again be obtained by the interchange of ${\mathrm Re}W_{sd}$ and 
${\mathrm Im}W_{sd}$ and multiplication of the right-hand-side 
by $\tau_{K_S} /\tau_{K_L}$.

\section{Numerical Calculation}

To obtain numerical predictions we have used a set of parameters 
taken from the Review of Particle Physics\cite{RPP99},
including the fine-structure constant at the weak scale, 
$\alpha = 1/129$;  $M_W = 80.3$ GeV; $\sin^2 \theta_W = 0.23$; 
and the measured semileptonic branching ratios needed in 
Eqs. (15), (16) and (18). We correspondingly find that:
\begin{eqnarray}
BR(K_L \to \pi^o \pi^o \nu \bar{\nu}) &=& 
[(3.1 \pm 0.6) ~\times 10^{-7} ] ~|{\rm Re}W_{ds}|^2 ~, \nonumber \\
BR(K_L \to (\pi^+ \pi^-)_{\rm I = 0} \nu \bar{\nu}) &=& 
[(6.2 \pm 1.2) ~\times 10^{-7} ] ~|{\rm Re}W_{ds}|^2 ~, \nonumber \\
BR(K_L \to (\pi^+ \pi^-)_{\rm I = 1} \nu \bar{\nu}) &=& 
[(0.93 \pm 0.05) ~\times 10^{-7} ] ~|{\rm Im}W_{ds}|^2 ~,
\end{eqnarray}
where the error bars come from those of the experimental measurements 
of the relevant semileptonic branching ratios. 
We have not taken account of radiative corrections or isotopic-spin 
violating differences in form factors and phase space, as has been 
done for the case of the decays involving a single pion\cite{MP96}, 
given the (larger) uncertainties in other parts of input at this 
stage of the analysis of these decays.  Formulas similar to Eq. (19) 
hold for the related decays of the $K_S$ to the same final states.

For the specific calculation of $W_{ds}$ in the Standard Model 
we need the values of $X(x_t )$ and of $X(x_c )$ in 
next-to-leading order\cite{BB93,BB99}, and that of the CKM matrix 
elements\cite{GKR99}
\begin{eqnarray}
{\mathrm Re} V_{td} &=& 0.0076 \pm 0.0015 ~,\nonumber \\
{\mathrm Im} V_{td} &=& 0.0031 \pm 0.0008 ~,
\end{eqnarray}
and $V_{ts} = - V_{cb} = - 0.040 \pm 0.002$, aside from the well-known 
matrix elements connecting the first and second generations. 
Using this and with $\bar{m_t} = 166 \pm 5$ GeV, we find that 
\begin{equation}
W_{ds}^{SM} = [(-6.7 \pm 1.0) + i(1.9 \pm 0.5 )] \times 10^{-4} ~,
\end{equation}
and the branching ratios for the various processes shown in Table I. 

In $K_L$ decays the contribution of the isospin one $\pi^+ \pi^-$ 
final state is negligible, since it is already suppressed compared 
to that with isospin zero from Eq.~(19) and the magnitude of the real 
part of  $W_{ds}^{SM}$ is considerably greater than that of the 
imaginary part.  Thus the ratio between $\pi^+ \pi^-$ and $\pi^o \pi^o$ 
rates is very close to the factor of two characteristic of isospin zero. 

\begin{table}
\begin{tabular}{ccc}
Processes & \multicolumn{2}{c}{Br($\times 10^{-13}$)} \\ \cline{2-3}
 & SM & New Physics (maximal) \\ \hline \hline
$K_L \to \pi^0 \pi^0 \nu \bar\nu$ & 
$1.4 \pm 0.4$ & $10$ \\
$K_S \to \pi^0 \pi^0 \nu \bar\nu$ & 
$(1.9 \pm 0.8) \times 10^{-4}$ & $8 \times 10^{-3}$ \\
$K_L \to \pi^+ \pi^- \nu \bar\nu$ & 
$2.8 \pm 0.8$ & $20$ \\
$K_S \to \pi^+ \pi^- \nu \bar\nu$ & 
$(11 \pm 2) \times 10^{-4}$ & $2 \times 10^{-2}$ \\
\end{tabular} \vspace{8pt}
\caption{Branching ratios of various $K_{L,S} \to \pi \pi \nu \bar\nu$
modes within and beyond the Standard Model.}
\end{table}

Our results are given in Table I, and both $K_L$ branching ratios lie 
between $10^{-13}$ and $10^{-12}$.  Our predictions in the Standard 
Model for $BR(K_L \to \pi^o \pi^o \nu \bar{\nu})$ are consistent with 
the previous calculation\cite{LV96} of $1 - 3 \times 10^{-13}$ and 
those\cite{GHL94,LV96} of  $1.1 - 5 \times 10^{-13}$ and 
$2 - 5 \times 10^{-13}$ for $BR(K_L \to \pi^+ \pi^- \nu \bar{\nu})$, 
but the allowed range is now considerably restricted. The $K_S$ branching 
ratios are in the $10^{-17}$ range in the Standard Model. The decay 
$K_S \to \pi^+ \pi^- \nu \bar{\nu}$ gets important contributions 
from both $I = 0$ and $I = 1$ $\pi \pi$ final states since the suppression 
of the $I = 1$ final state is compensated by the ratio of 
$({\mathrm Re} W_{ds}/{\mathrm Im} W_{ds})^2$ in the Standard Model.

For the representative examples of physics beyond the Standard Model, 
we also show in Table I values for the branching ratios that correspond 
to the maximal values one could obtain consistent with the bounds 
in Eqs. (6)- (9) and (12) - (13), respectively.  These maximal values 
are similar in both cases and arise when the parameters of the new 
physics are chosen to maximize $W_{ds}^{NP}$ consistent with the 
constraints coming from known K physics. Among these constraints is the 
recent branching ratio measurement for $K^+ \to \pi^+ \nu \bar\nu$, 
which is equally sensitive to both the real and imaginary parts 
of the total $W_{ds}$ with minimal theoretical assumptions.  
Hence, similar maximal values of $W_{ds}^{NP}$ are obtained in any 
model of new physics.  Note that when taking ratios to the 
Standard Model branching ratios, a much bigger 
factor is possible when the new physics enters the imaginary 
part of $W_{sd}^{NP}$, and is CP-violating, since the imaginary 
part of $W_{sd}^{SM}$ is considerably smaller than the real part. 

\section{Summary}

We have used recent information on the CKM matrix to narrow the 
range of the predicted branching ratios for  $K_{L,S} \to \pi^o \pi^o 
\nu \bar{\nu}$ and $K_{L,S} \to \pi^+ \pi^- \nu \bar{\nu}$ decays 
in the Standard Model.  These branching ratios are in the neighborhood 
of $ 1~-~4 \times 10^{-13}$ for the $K_L$ decays, which make them 
possibly observable at the few event level in the next round of 
experiments that are setting out to see the CP-violating decays 
with a single $\pi^o$ in the final state.  These branching ratios 
could be larger by up to about an order of magnitude in theories 
that go beyond the Standard Model.

The branching ratios for the $K_S$ decays are in the neighborhood 
of $10^{-17}$ to $10^{-16}$ in the Standard Model and seem 
unlikely to ever be observed.  Here new physics could boost the 
branching ratios by more than an order of magnitude, although 
even then the maximum branching ratio of around $10^{-15}$ is still 
beyond the limits of observation.   For both $K_S$ and $K_L$ decays, 
increased experimental accuracy in the measurement of the branching 
ratio for $K^+ \to \pi^+ \nu \bar{\nu}$, 
assuming it remains consistent with the Standard Model, will put 
more stringent restrictions on non-Standard-Model physics in the 
$s \to d \nu \bar{\nu}$ transition and limit the deviations from 
the Standard Model that can be observed in the decays under 
discussion here as well.

\noindent{\bf ACKNOWLEDGMENT}

This research work is supported by the Department of Energy under
Grant No. DE-FG02-91ER40682.  Fred Gilman thanks Y. Wah for 
discussions that started this investigation.

{\tighten

}

\end{document}